\documentclass[journal]{IEEEtran}

\usepackage{amsmath, amssymb, bm, cite, epsfig, psfrag}
\usepackage{epstopdf}
\usepackage{graphicx}
\usepackage{algorithm,algpseudocode}
\usepackage{array}
\usepackage[margin=10pt,font=small]{caption}
\usepackage[margin=.71in,bottom=1in]{geometry}
\usepackage{bbm}
\usepackage{multirow}
\usepackage[usenames,dvipsnames]{xcolor}

\usepackage{subfigure}
\usepackage{etoolbox}
\usepackage{pbox}
\usepackage[width=0.48\textwidth]{caption}
\usepackage{colortbl}
\usepackage{fancyhdr}
\newtoggle{conference}
\togglefalse{conference} 
\graphicspath{{figures/}}

\def\beq{\begin{equation}}
\def\eeq{\end{equation}}
\def\beqa{\begin{eqnarray}}
\def\eeqa{\end{eqnarray}}
\def\beqan{\begin{eqnarray*}}
\def\eeqan{\end{eqnarray*}}

\def\tm1{t\! - \! 1}
\def\tp1{t\! + \! 1}

\def\BW{\mathrm{BW}}

\usepackage{lipsum}
\usepackage{fancyhdr}
\usepackage{datetime}

\usepackage{tikz}
\usetikzlibrary{calc}

\usepackage{longtable}

\begin{document}

\newcommand\blfootnote[1]{%
  \begingroup
  \renewcommand\thefootnote{}\footnote{#1}%
  \addtocounter{footnote}{-1}%
  \endgroup
}

\pagestyle{empty}

\title{MIMO Channel Modeling and Capacity Analysis for 5G Millimeter-Wave Wireless Systems}

\author{
	\IEEEauthorblockN{Mathew K. Samimi, Shu Sun, and Theodore S. Rappaport\\NYU WIRELESS, NYU Tandon School of Engineering\\mks@nyu.edu, ss7152@nyu.edu, tsr@nyu.edu} \\

\thanks{The authors wish to thank the NYU WIRELESS Industrial Affiliates for their support, and G. R. MacCartney Jr., S. Sun, S. Deng, T. Wu, M. Zhang, J. P. Ryan,  A. Rajarshi for their contribution, and Prof. S. Rangan for valuable discussions. This work is supported by National Science Foundation (NSF) Grants (1302336, 1320472, and 1555332).}
}

\maketitle

\begin{tikzpicture} [remember picture, overlay]
\node at ($(current page.north) + (0,-0.25in)$) {M. K. Samimi, S. Sun, T. S. Rappaport, ``MIMO Channel Modeling and Capacity Analysis for 5G Millimeter-Wave Wireless Systems,''};
\node at ($(current page.north) + (0,-0.4in)$) {\textit{in the 10\textsuperscript{th} European Conference on Antennas and Propagation (EuCAP'2016)}, April 2016.};
\end{tikzpicture}

\begin{abstract}
This paper presents a 3-D statistical channel model of the impulse response with small-scale spatially correlated random coefficients for multi-element transmitter and receiver antenna arrays, derived using the physically-based \textit{time cluster - spatial lobe} (TCSL) clustering scheme. The small-scale properties of multipath amplitudes are modeled based on 28 GHz outdoor millimeter-wave small-scale local area channel measurements. The wideband channel capacity is evaluated by considering measurement-based Rician-distributed voltage amplitudes, and the spatial autocorrelation of multipath amplitudes for each pair of transmitter and receiver antenna elements.  Results indicate that  Rician channels may exhibit equal or possibly greater capacity compared to Rayleigh channels, depending on the number of antennas. 
\end{abstract}

\section{Introduction}

A rich multipath fading environment can be effectively utilized in multiple-input multiple-output (MIMO) wireless systems to dramatically increase system capacity~\cite{Pau04}, by simultaneously exploiting independent spatial subchannels between transmitter (TX) and receiver (RX) antenna elements. The MIMO channel capacity is limited by the number of antennas, the antenna element spacing, and the transmit and receiver spatial correlations, requiring realistic models to estimate channel coefficients and total channel capacity for MIMO system-level design. At millimeter-wave (mmWave) frequencies, hundreds of electrically-steerable antennas can be placed on a transceiver to compensate for free space path loss and enable beamforming algorithms that support MIMO communications~\cite{Rap15}.

Previous work has investigated the MIMO channel capacity for a wideband (i.e., frequency-selective) channel to enable channel modeling for the design of broadband systems and very high data rates. The widespread 3GPP and WINNER 3-D spatial channel models (SCM) are suitable for MIMO system-level designs below 6 GHz, and use a cluster-level modeling approach in which the spatially fading MIMO channel coefficients are obtained from the superposition of cluster subpath powers across antenna array elements. However, note that small-scale spatial fading distributions and spatial autocorrelation models are not specified to simulate local area effects~\cite{3GPP:1,WinnerII}. The effects of spatial and temporal correlations of multipath amplitudes at different antenna elements affect MIMO capacity results, and must be appropriately modeled from measurements to enable realistic multi-element antenna simulations. 

Work in~\cite{Shiu00} demonstrates the importance of spatial and temporal fading correlations in a MIMO communication system, showing a large increase in capacity over single-input single-output (SISO) systems when the fades connecting pairs of TX and RX antennas are independent and identically distributed (i.i.d.). In~\cite{Molisch02,Inta04}, the spatial correlations of transmitter and receiver arrays are reconstructed from measured AODs and AOAs of the multipaths. The correlation-based models are used to simulate the MIMO system capacity, which is found to agree relatively well with the measured MIMO capacity.

Knowledge of the spatial correlations of multipath amplitudes at the TX and RX enable the design of antenna arrays to maximize channel capacity~\cite{Inta04}. In~\cite{Byers04}, a space-time cross-correlation function is derived from measurements, that describes the joint spatial-temporal correlations in a MIMO system for a mobile scenario. Outage capacities were found to increase linearly with the number of antennas~\cite{Byers04}. Various schemes may be implemented on a MIMO radio channel to further enhance system capacity, for instance, antenna selection~\cite{Li12}.

In this paper, the mmWave SISO modeling approach derived from the \textit{time cluster - spatial lobe} (TCSL) clustering algorithm~\cite{Samimi15_3} is extended to a MIMO channel model for arbitrary antenna pattern using measurement-based spatial autocorrelation functions and small-scale spatial fading distributions of multipath amplitudes~\cite{Samimi16}, to generate power delay profiles (PDPs) over a local area. The system capacity of a MIMO system at 28 GHz is then investigated using Monte Carlo simulations in a realistic mmWave MIMO mobile radio channel based on the statistical spatial channel models (SSCM) in~\cite{Samimi15_3,Samimi16}, to enable next generation mmWave air interface design~\cite{Fung93}.

\section{3-D Local Area Channel Impulse Response}
\label{secII}
The omnidirectional radio propagation channel can be described using the double-directional time-invariant baseband channel impulse response~\cite{Steinbauer01}, also known as a parametric channel model~\cite{For07} and commonly expressed as in~(\ref{eq1}),
\begin{equation}\label{eq1}
\begin{split}
h(t,\overrightarrow{\mathbf{\Theta}},\overrightarrow{\mathbf{\theta}}) &= \sum_{k=1}^K a_k e^{j\varphi_k} \delta(t - \tau_k)\\
& \cdot \delta(\overrightarrow{\mathbf{\Theta}}-\overrightarrow{\mathbf{\Theta}_k}) \cdot \delta(\overrightarrow{\mathbf{\Phi}}-\overrightarrow{\mathbf{\Phi}_k})
\end{split}
\end{equation}

\noindent where $a_k$, $\varphi_k$, and $\tau_k$ are the path voltage amplitude, phase, and absolute propagation delay of the $k$\textsuperscript{th} multipath component (MPC); $\overrightarrow{\mathbf{\Theta}_k}$ and $\overrightarrow{\mathbf{\Phi}_k}$ are the vectors of azimuth/elevation angle of departure (AOD) and angle of arrival (AOA), respectively; $K$ is the total number of multipath components. Statistical distributions for $|a_k|^2$, $\varphi_k$, $\tau_k$, $\overrightarrow{\mathbf{\Theta}_k}$, $\overrightarrow{\mathbf{\Phi}_k}$, and $K$ shown in~(\ref{eq1}) are derived in~\cite{Samimi15_3} using the time cluster - spatial lobe (TCSL) algorithm, in which a physically-based clustering method is applied separately in the delay and angular domains to extract channel parameters~\cite{Samimi15_2,Samimi15_3}. 

Time clusters represent groups of multipath components arriving within a short propagation time window, but that can arrive from potentially widely varying angles of arrival. A spatial lobe denotes a direction of arrival (or departure) where energy is contiguous over the azimuth and elevation dimensions, and where multipaths can arrive over many hundreds of nanoseconds. The statistics of time clusters and spatial lobes are extracted separately from channel measurements, by applying a 25 ns minimum inter-cluster void interval used to partition multipath time of arrivals~\cite{Samimi15_3}, and by thresholding the power angular spectrum (PAS) based on a -10 dB and -20 dB lobe power threshold with respect to the maximum received power in the PAS. The TCSL algorithm has previously been shown to recreate first- and second-order delay and angular statistics~\cite{Samimi15_3}.

Nonparametric channel models~\cite{For07} are commonly used to describe the stochastic evolution of the MIMO channel matrix $\textbf{H}_l$, where $\textbf{H}_l$ denotes the $\textit{N}_r \times \textit{N}_t$ MIMO channel matrix of the $l$\textsuperscript{th} multipath component in an omnidirectional channel impulse response, expressed as in~(\ref{eq:Hl})~\cite{For07}:
\begin{equation}\label{eq:Hl}
\textbf{H}_l=\textbf{R}_r^{1/2}\textbf{H}_w\textbf{R}_t^{1/2}
\end{equation}

\noindent where $\textbf{R}_r$ and $\textbf{R}_t$ denote the receive and transmit spatial correlation matrices, respectively, for user-defined antenna array, and $\textbf{H}_w$ is a matrix whose entries correspond to small-scale spatial path (voltage) amplitudes and phases. Note that $\textbf{R}_r$ and $\textbf{R}_t$ collapse to the identity matrices when disregarding the spatial correlation of multipath across the antenna elements. The entries of $\textbf{H}_w$ are commonly assumed i.i.d. Rician and Rayleigh in line of sight (LOS) and non-line of sight (NLOS) environments, respectively. The entries of $\textbf{H}_l$ retain the spatial autocorrelation of multipath amplitudes specified through $\textbf{R}_r$ and $\textbf{R}_t$, while exhibiting the small-scale distribution specified in  $\textbf{H}_w$. 

The entries of $\textbf{R}_t$ and $\textbf{R}_r$ can be obtained from measurement-based spatial autocorrelation of multipath amplitudes~\cite{Rap91,Karttunen98}, or directly from measured AOD and AOA power angular spectra~\cite{Inta04,Molisch02}. In this paper, the empirical spatial autocorrelation functions of multipath amplitudes are provided and used to generate the spatial correlation matrices.

The wideband local area channel impulse response (CIR) and corresponding MIMO channel capacity can be obtained as follows:
\begin{enumerate}
\item Generate one initial CIR using~(\ref{eq1}), from the mmWave SISO channel model in~\cite{Samimi15_3}. This is the mean spatially averaged CIR.
\item For each generated multipath component in the initial CIR, generate $N_r\times N_t$ local ``copies'' over the local area, using~\eqref{eq:Hl}, such that the voltage magnitudes obey the spatial correlation specified by $\textbf{R}_r$ and $\textbf{R}_t$ and the small-scale distribution specified by $\textbf{H}_w$. The multipath delays, AODs, and AOAs of each multipath copy over the local area remain identical to the delays and angles in the initial CIR.
\item Compute the frequency response $\textbf{H}_f$ of the MIMO channel impulse response $\textbf{H}_l$ using a discrete Fourier transform operation.
\item Compute the total wideband capacity from~(\ref{eqCap})~\cite{Molisch02}:
\begin{equation}\label{eqCap}
C=\frac{1}{\BW}\int_{f_{min}}^{f_{max}} \log_2\det\left(\textbf{I}+\frac{\rho}{N_t}\textbf{H}_f\textbf{H}_f^H\right)df
\end{equation}

\noindent where $\BW$ denotes bandwidth, $\rho$ represents the average SNR, $f_{min}$ and $f_{max}$ denote the minimum and maximum narrowband sub-carrier frequencies, respectively.
\end{enumerate}

\section{Small-Scale Measurement Descriptions}

The 28 GHz small-scale track measurements were performed using a 400 megachips-per-second (Mpcs) broadband sliding correlator channel sounder, and a pair of high gain 15 dBi directional horn antennas ($28.8^{\circ}$ and $30^{\circ}$ half-power beamwidths in azimuth and elevation, respectively) at one TX and four RX locations~\cite{Samimi16}. The distance between the transmitter and the center of the RX local areas ranged from 8 m to 12.9 m. The maximum measurable path loss was 157 dB, with a measurement time resolution of 2.5 ns (800 MHz RF null-to-null). At each RX location, the RX antenna was moved over a 33-wavelength long track, emulating a virtual array with antenna spacing of $\lambda/2=5.35$ mm, where each antenna position was situated on a cross (i.e., 66 antenna positions on each axis of the cross). The TX and RX antennas were fixed in azimuth and elevation while PDPs were acquired for each step increment with fixed RX antenna during the captures. The RX and TX antennas were located 1.4 m and 4 m above ground level, respectively, well below surrounding rooftops. Directional antennas were employed to emulate a typical realistic mmWave base-to-mobile scenario, where both the TX and RX beamform towards the strongest angular directions.
\section{Measurement-Based Statistical Models}

\subsection{Millimeter-Wave Small-Scale Spatial Fading}

\begin{figure}
    \centering
 \includegraphics[width=3.5in]{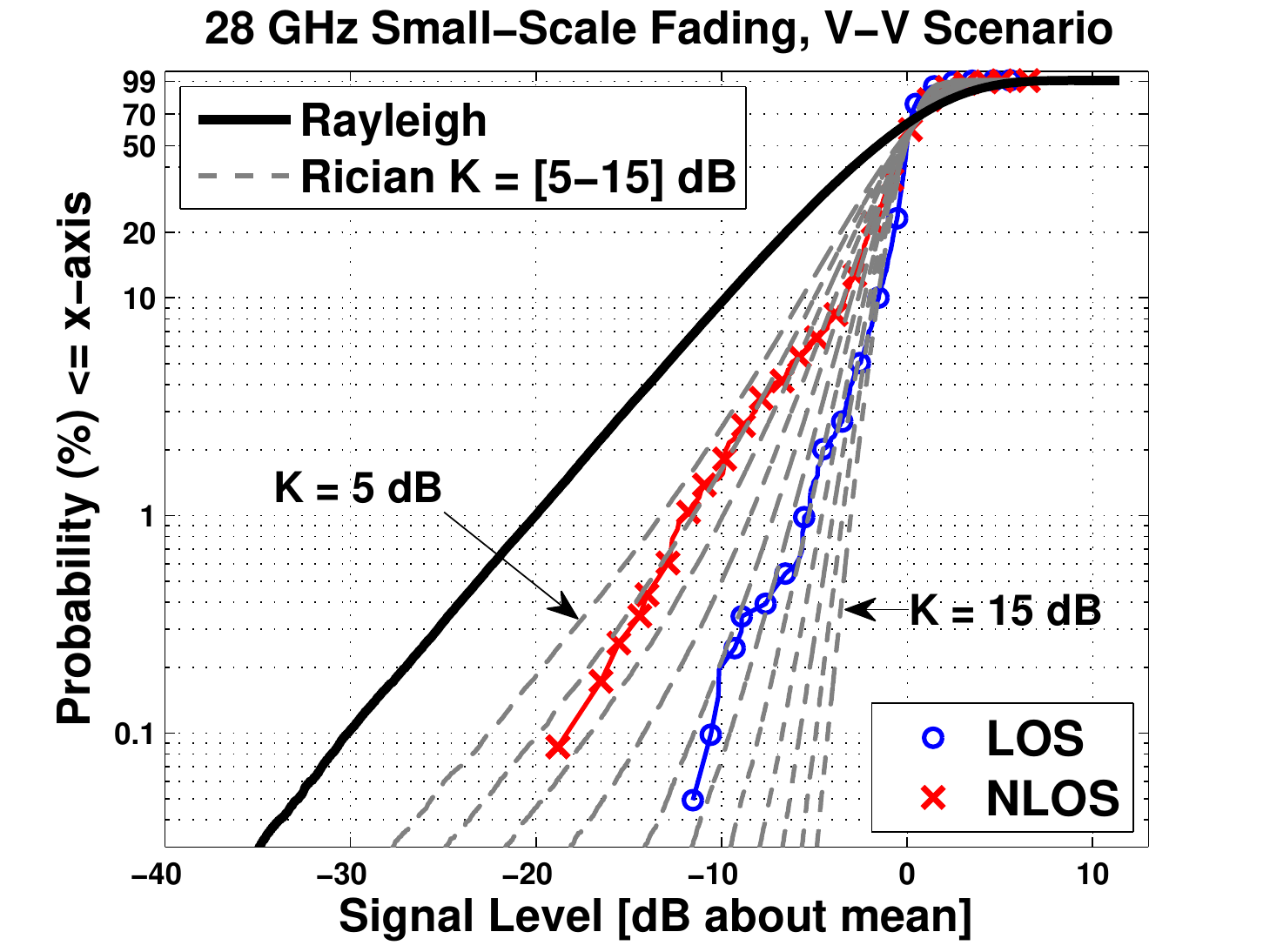}
    \caption{CDF of 28 GHz measurement-based small-scale spatial fading distributions in LOS and NLOS scenarios.}
    \label{fig:F1}
\end{figure}

Small-scale fading denotes the fluctuations in received signal levels over short, sub-wavelength receiver distances, and is physically explained by the coherent phasor sum of many random multipath components arriving within the measurement system resolution~\cite{Rap02}. Fig.~\ref{fig:F1} shows the cumulative distribution functions (CDFs) for $|a_k|^2/\overline{|a_k^2|}$ in LOS and NLOS, superimposed with a Rayleigh distribution, and Rician distributions plotted for various $K$ factors ranging from 5 dB to 15 dB, in steps of 1 dB~\cite{Samimi16}. The small-scale fading distributions tend to follow a Rician distribution, compared to the traditional Rayleigh distribution, indicating the presence of a strong dominant path and a few weak scattered multipaths~\cite{Samimi16}. The Rician distribution fit all measured data, in both LOS, NLOS and LOS-to-NLOS environments, for the V-V and V-H scenarios investigated.
Table~\ref{tbl:T1} summarizes the various $K$ factors as a function of environment and polarization configuration.

\begin{table}
\centering
\caption{Table summarizing the ranges of $K$-factors for the Rician distributions, describing the path (voltage) gains $a_k$ in~(\ref{eq1}), obtained from 28 GHz directional small-scale fading measurements over a local area in different environments, for V-V and V-H polarization configurations.}

\begin{tabular}{|c|c|c|}

\hline
\textbf{Environment}	& \bm{$K_{VV}} \textbf{ [dB]}$	& \bm{$K_{VH}} \textbf{ [dB]}$ \\ \hline

LOS				& 9 - 15					& 3 - 7				\\ \hline
NLOS				& 5 - 8					& 3 - 7				\\ \hline
LOS-to-NLOS			& 4 - 7 					& 6 - 10				\\ \hline
\end{tabular}
\label{tbl:T1}
\end{table}

\subsection{Average Spatial Autocorrelation of Multipath Amplitudes}
\begin{figure}
    \centering
 \includegraphics[width=3.5in]{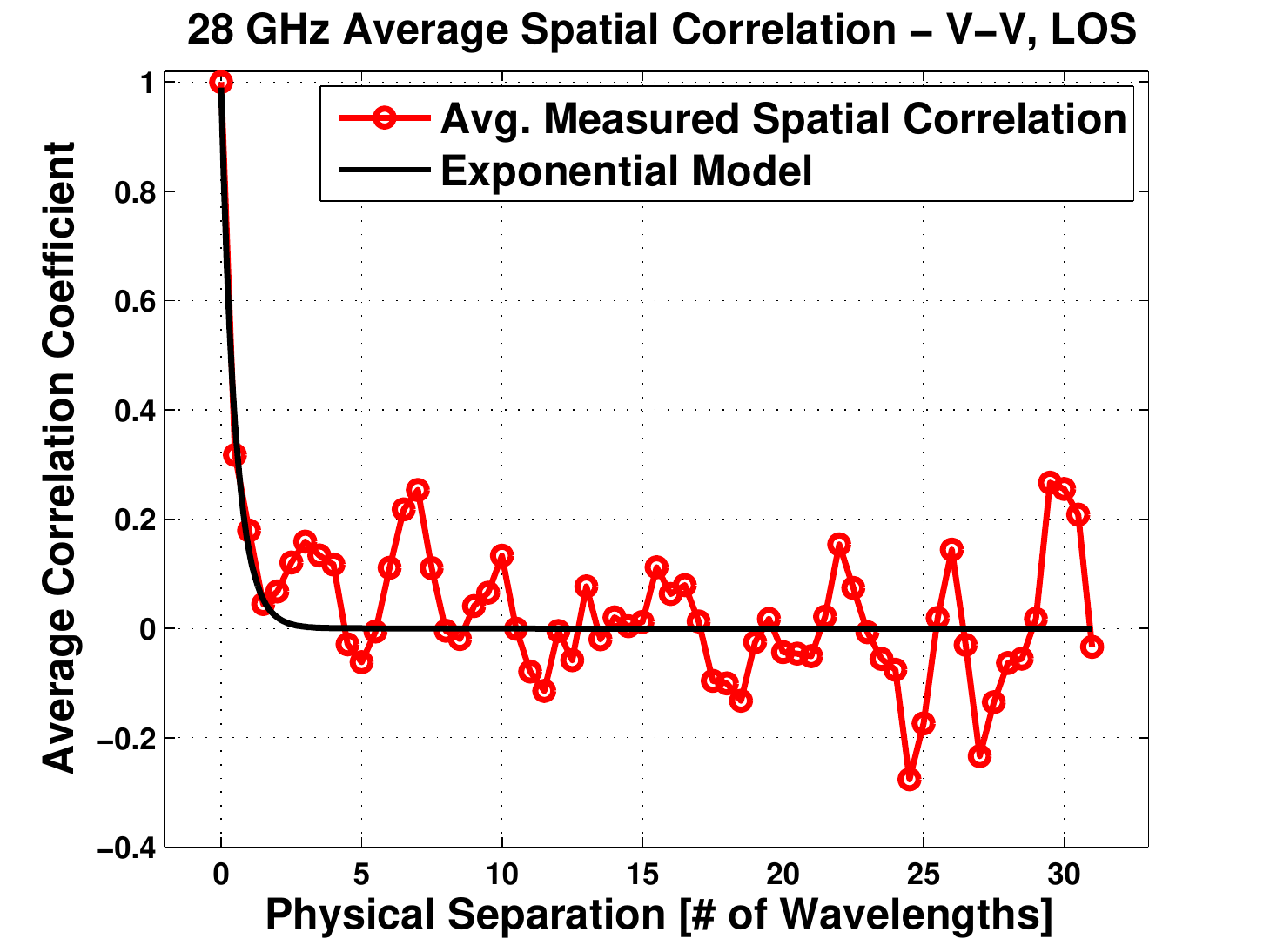}
    \caption{Empirical average spatial autocorrelation of resolvable multipath amplitudes, and mean exponential model obtained at 28 GHz vertical-to-vertical LOS scenario.}
    \label{fig:F2}
\end{figure}

\begin{figure}
    \centering
 \includegraphics[width=3.5in]{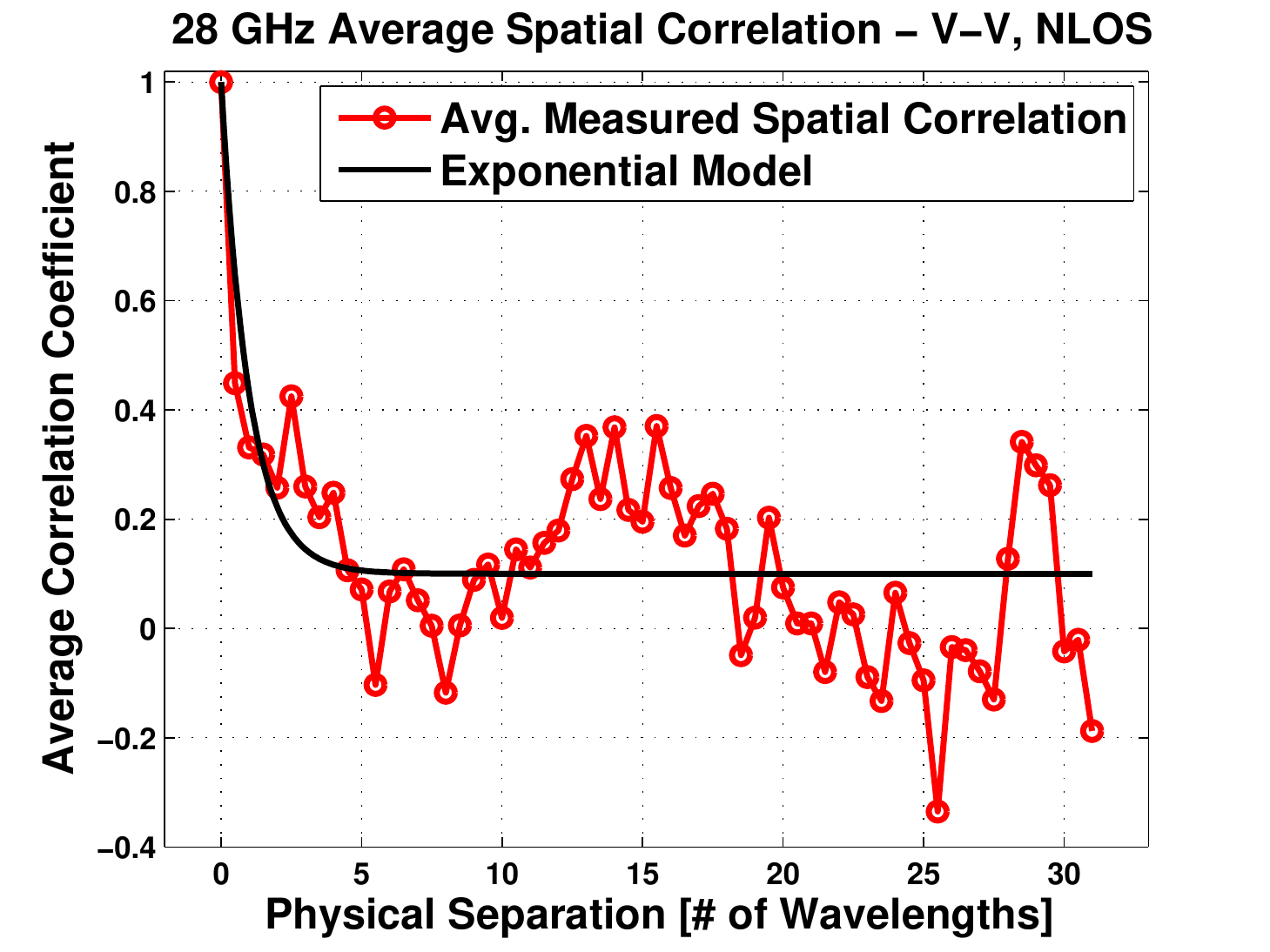}
    \caption{Empirical average spatial autocorrelation of resolvable multipath amplitudes, and mean exponential model obtained at 28 GHz vertical-to-vertical NLOS scenario.}
    \label{fig:F3}
\end{figure}

The spatial autocorrelation of individual multipath component voltage amplitudes indicates the level of similarity in signal levels between antennas $i$ and $j$ separated by $\Delta r$. The spatial autocorrelation values were computed from~(\ref{Auto}) using all co-polarized and cross-polarized measurements in LOS and NLOS environments, where E[] is the expectation operator over the ensemble data, $\Delta X$ is the physical separation between two adjacent track positions, and is equal to $\lambda / 2 = 5.35$ mm, $A_K(T_K,X_l )$ is the multipath voltage amplitude at track position $l$ and bin delay $T_K$~\cite{Rap91}. Fig.~\ref{fig:F2} and Fig.~\ref{fig:F3} show the average (over excess delay) spatial autocorrelation function obtained from the V-V measurements in LOS and NLOS scenarios, and the corresponding best fit exponential model of the form~\cite{Karttunen98},
\begin{equation}\label{eq3}
f(\Delta r) = A e^{-B \Delta r}-C\end{equation}

\noindent where $A$, $B$, and $C$ are constants that were determined using the minimum mean square error (MMSE) method, by minimizing the error between the empirical curve and theoretical exponential model shown in~(\ref{eq3}). In Fig.~\ref{fig:F2} and Fig.~\ref{fig:F3}, the constants were determined to be $A=0.99$, $B=1.95$, $C=0$, and $A=0.9$, $B=1$, $C=-0.1$, respectively. Table~\ref{tbl:T2} summarizes the model coefficients as a function of polarization and environment type.

\begin{figure*}
\begin{equation}\label{Auto}
\rho(i \Delta X) = \frac{E \big[ \big( A_K (T_K,X_l)- \overline{A_K(T_K,X_l)} \big)    \big( A_K (T_K,X_l+i\Delta X)- \overline{A_K(T_K,X_l +i \Delta X )} \big)       \big] }{\sqrt{E \bigg[ \big( A_K (T_K,X_l)- \overline{A_K(T_K,X_l)} \big)^2  \bigg] E\bigg[ \big( A_K (T_K,X_l+ i \Delta X)- \overline{A_K(T_K,X_l + i \Delta X )} \big)^2       \bigg]}}, i = 0, 1, 2, ...
\end{equation}
\hrulefill
\vspace*{4pt}
\end{figure*}

\begin{table}
\centering
\caption{Table summarizing the model parameters (A, B, C) in~(\ref{eq3}) obtained using the MMSE method, to estimate the empirical spatial autocorrelation functions.}

\begin{tabular}{|c|c|c|}

\hline
\textbf{(A, B, C)}		& \textbf{V-V}				& \textbf{V-H}				 \\ \hline

LOS				& (0.99, 1.95, 0)				& (1.0, 0.9, 0.05) 				\\ \hline
NLOS				& (0.9, 1, -0.1)				& (1, 2.6, 0)				\\ \hline
\end{tabular}
\label{tbl:T2}
\end{table}

\section{Simulations}
\subsection{Simulation Settings}
A Monte Carlo simulation was performed to simulate the wideband capacity using a realistic mmWave measurement-based single-input multiple-output (SIMO) and MIMO channel model. The carrier frequency is centered at 28 GHz with a bandwidth of 800 MHz, which is uniformly divided into 100 narrowband sub-carriers. For spatially correlated small-scale distribution in a ULA, the spatial correlation matrices are calculated using~\eqref{eq:rxCov_ULA3}:
\begin{equation}\label{eq:rxCov_ULA3}
\begin{split}
\left[\textbf{R}_{r}\right]_{i,k}&=e^{-j\Theta}(A e^{-B|i-k|d}-C)
\end{split}
\end{equation}

\noindent where $\Theta$ is a random phase assigned to each coefficient $\left[\textbf{R}_{r}\right]_{i,k}$, and $\Theta=0$ when $i=k$. For the Rician distribution using~\eqref{eq:rxCov_ULA3}, the parameter values corresponding to the NLOS V-V scenario ($A=0.9$, $B=1$, $C=-0.1$) are adopted.  

A ULA with 20 antenna elements with a spacing of $\lambda / 2$ is simulated at the receiver, while a ULA with one antenna element is used at the transmitter, forming a SIMO channel. Next, the number of transmit antenna elements $N_t$ is set to 2 to form a MIMO channel together with the 20 receive antenna elements in the ULA. Capacity comparisons are made between different small-scale spatial distributions, including the Rician distribution derived from Fig.~\ref{fig:F1}, and the widely-used Rayleigh distribution assumption. 

\begin{figure}
    \centering
 \includegraphics[width=3.5in]{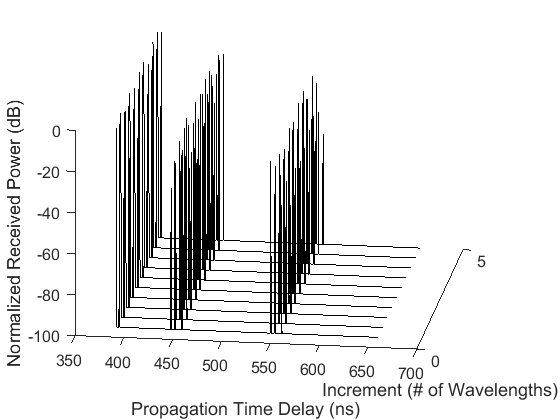}
    \caption{Output of the SIMO local area PDPs sampled every half-wavelength over a linear track of five wavelengths in a vertical-to-vertical NLOS scenario, where the small scale spatial distribution follows a Rician distribution with a $K$ factor of 5 dB, and the spatial autocorrelation obeys Eq.~\eqref{eq:rxCov_ULA3}.}
    \label{fig:PDPs}
\end{figure}

Fig.~\ref{fig:PDPs} illustrates the sample output of a SIMO local area PDP sampled every half-wavelength over a linear track of five wavelengths in a vertical-to-vertical NLOS scenario using the four steps outlined in Section~\ref{secII}, where the small scale spatial coefficients follow a Rician distribution with a $K$ factor of 5 dB, and the spatial autocorrelation of multipath amplitudes obeys Eq.~\eqref{eq:rxCov_ULA3}. In Fig.~\ref{fig:PDPs}, the amplitudes of the multipath component at the same time delay remain relatively constant over the local area, as observed from the measurements. 

\subsection{Simulation Results}
\begin{figure}
    \centering
 \includegraphics[width=3.5in]{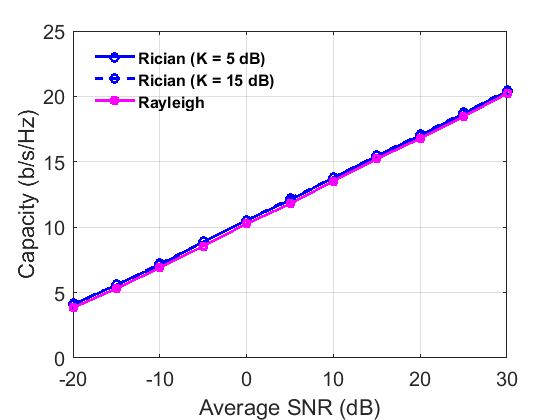}
    \caption{Comparison of SIMO channel capacity between Rayleigh and Rician distributed small-scale fading coefficients with Rician $K$ factors of 5 dB and 15 dB at 28 GHz. A single antenna is used at the transmitter, and a ULA with 20 elements is used at the receiver.}
    \label{fig:ULA_RC_RL}
\end{figure}

Fig.~\ref{fig:ULA_RC_RL} compares the SIMO channel capacity using Rayleigh and Rician distributed small-scale spatial fading coefficients with Rician $K$ factors of 5 dB and 15 dB, using Eq.~\eqref{eqCap}. The Rician distribution yields slight improvement (about 0.3 b/s/Hz) in capacity compared to the Rayleigh distribution. The physical interpretation is that the smaller the number of antennas, the fewer the number of spatial streams, thus a strong component will contribute significant channel capacity; when there are many antennas, we can exploit the MPCs to a large extent, hence the more MPCs the larger the capacity.

\begin{figure}
    \centering
 \includegraphics[width=3.5in]{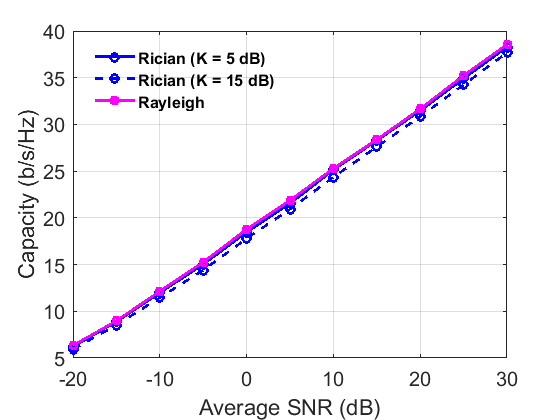}
    \caption{Comparison of MIMO channel capacity between Rayleigh and Rician distributed small-scale fading coefficients with Rician $K$ factors of 5 dB and 15 dB at 28 GHz. ULAs with 2 and 20 elements are used at the transmitter and receiver, respectively.}
    \label{fig:ULA_RC_RL_MIMO}
\end{figure}

 Fig.~\ref{fig:ULA_RC_RL_MIMO} compares the MIMO channel capacity between Rayleigh and Rician small-scale distributions with Rician $K$ factors of 5 dB and 15 dB at 28 GHz. It is observed that the Rayleigh distribution yields the highest capacity, while the lowest capacity is associated with the Rician distribution with a $K$ factor of 15 dB, i.e., the lower the $K$ factor, the higher the capacity in this case. These results are quite different from those shown by Fig.~\ref{fig:ULA_RC_RL}, which indicates that the Rician distribution may increase channel capacity, as confirmed by the analysis in~\cite{Kha01}. Rician channels may result in greater or smaller capacity compared to Rayleigh channels, depending on the number of antennas due to the compromise between multipath exploitation and fading reduction~\cite{Kha01}.
 
\section{Conclusion}
This paper presented a 3-D statistical channel model to simulate 28 GHz local area CIRs, derived from the TCSL clustering algorithm, suitable for MIMO system-level simulations. The small-scale spatial channel coefficients are Rician-distributed, with exponential spatial autocorrelation of multipath amplitudes. Further, the SIMO channel matrix has been extended to the MIMO case for arbitrary antenna arrays, but shown here for ULAs at the transmitter and receiver. Monte Carlo simulations evaluating wideband capacities indicate that the Rician distribution and the measurement-based spatial autocorrelation of multipath component amplitudes can yield higher channel capacity compared to the Rayleigh assumption, when the number of transmit antennas is small. 

\bibliographystyle{IEEEtran}
\bibliography{bibliography_MSThesis}
\end{document}